\documentclass[aps,prl,twocolumn,psfig,groupedaddress,floats,showpacs]{revtex4}

\usepackage{bm}

\usepackage{amssymb,amsmath}
\usepackage{latexsym}
\usepackage{graphicx}
\usepackage{epsfig}

\begin{document}

\preprint{APS/123-QED}

\title{Femtosecond Coherent Control of Spin with  Light in 
(Ga,Mn)As ferromagnets 
}

\author{M. D. Kapetanakis and I. E. Perakis}

\affiliation{Department of Physics, 
University of Crete, Heraklion, Crete, 71003
and 
Institute of Electronic Structure \& Laser, Foundation
for Research and Technology-Hellas, Heraklion, Crete, 
71110, Greece}

\author{K. J. Wickey and  C. Piermarocchi}

\affiliation{Department of Physics 
\& Astronomy,
 Michigan State University, East Lansing, MI, 48824, USA}

\author{J. Wang}

\affiliation{Department of Physics 
\& Astronomy and Ames Laboratory-USDOE, 
Iowa State University, Ames, IA 50011, USA.}

\date{\today}

\begin{abstract} 
Using  density matrix equations of motion,  
we predict  
a femtosecond 
 collective spin tilt triggered by 
nonlinear, near--ultraviolet ($\sim$3eV), coherent 
photoexcitation of  (Ga,Mn)As ferromagnetic semiconductors  
with linearly polarized light. 
This dynamics 
results from  
carrier coherences 
and nonthermal populations
excited in the \{111\}
equivalent directions of the Brillouin zone
and triggers a subsequent uniform  precession.
We predict nonthermal magnetization 
control 
by  tuning the 
laser frequency and polarization direction.
Our  mechanism explains recent 
ultrafast  pump--probe experiments.  
\end{abstract}
\pacs{78.47.J-, 78.20.Ls, 78.47.Fg, 42.50.Md }

\maketitle

Long range magnetic order 
arises from the interactions between itinerant and localized 
spins
in a wide variety  of systems, 
such as EuO, EuS, 
chrome spinels,  pyrochlore, 
manganese oxides, or 
(III,Mn)V ferromagnetic semiconductors \cite{nagaev,rmp}. 
With ferromagnetic semiconductors 
one can envision multifunctional devices combining
information processing and  storage
on a single chip with low power consumption. 
Fast spin manipulation is of great importance  for such  
spin--electronic,
spin--photonic, 
magnetic storage, 
and quantum computation applications.

One of the challenges facing magnetic 
devices concerns their speed.
The  magnetic properties of carrier--induced ferromagnets 
respond strongly to 
carrier density tuning via light, 
electrical gates, or current 
\cite{ohno2000}. 
While magnetic field pulses and spin currents can be  
used to manipulate spin on the many--picosecond 
time scale, 
femtosecond spin manipulation 
requires 
the use of laser pulses \cite{kimel,wang}. 
In ultrafast pump--probe magneto--optical spectroscopy, 
the  pump optical pulse 
excites 
e--h coherences 
and 
corresponding
carrier populations, whose subsequent interactions 
trigger a magnetization dynamics
monitored as function of time
via the 
Faraday or Kerr rotation  \cite{demag}.

The physical processes  leading to  
femtosecond  magnetization dynamics 
(femto--magnetism)
are   under debate. 
Open questions include the possibility of direct 
photon--spin coupling, the 
distinction of coherent and incoherent effects, and the 
exact role of the spin--orbit interaction.  
Following the pioneering work of Ref.\cite{bigot1}, 
many ultrafast spectroscopy experiments
were interpreted in terms of 
a decrease in the  magnetization {\em amplitude}
due to transient {\em thermal} effects 
\cite{bigot1,wang05}.
Observations of light--induced changes in the magnetization 
{\em orientation} 
were also mostly attributed to the 
temperature elevation, which leads to 
transient changes 
in the magnetic easy axes
\cite{bigot-05,tolk,merlin}.
Most desirable is {\em nonthermal} magnetization control 
within  
the femtosecond coherent \cite{sbe} 
temporal regime, which 
promises more flexibility limited only by the optical 
pulse duration. Experiments in 
ferrimagnetic garnets were interpreted in terms of 
an interplay between the inverse Faraday effect
\cite{farad}
and long--lived changes in the magneto--crystalline anisotropy 
\cite{kimel}. 
In (Ga,Mn)As, Ref.\cite{hash}
reported  magnetization  precession 
triggered by 
changes of magnetic anisotropy 
on a $\sim$100ps
 time scale
due to carrier relaxation, 
while Ref.\cite{rozk}  
demonstrated coherent control of the
precession. 
Recently, Wang {\it et.al} \cite{wang} 
observed 
{\em two} distinct temporal 
regimes of magnetization 
re-orientation 
when  (Ga,Mn)As is  
excited by  3.1eV photons. 
Prior to the 
$\sim$100ps  precession,  
they observed a
 {\em quasi--instantaneous}  
magnetization tilt, which was absent for
excitation 
 near the GaAs fundamental gap
($\sim$1.55eV).  

In this letter we calculate 
the Mn 
spin dynamics triggered by 
femtosecond coherent photoexcitation of  (Ga,Mn)As 
with linearly polarized light.
The joint density of states for interband transitions 
has a strong peak in the neighborhood of 3eV, due to $\Lambda_3
\rightarrow \Lambda_1$ excitations along the eight equivalent
directions $\{111\}$ of the Brillouin zone (BZ), the $\Lambda$--edge
\cite{burch}. 
 Using a
full tight-binding calculation of the bands, we show that
photoexcitation at the $\Lambda$--edge is advantageous for
non--thermal spin manipulation.  This is due to the interplay between
BZ symmetry, spin--orbit
and magnetic exchange interactions,  
and coherent nonlinear photoexcitation. 
We predict a femtosecond 
 temporal regime 
dominated by coherent e--h pairs and nonthermal 
populations. This is followed by a second  regime 
determined by the magnetic anisotropy 
of the $\Gamma$--point hole Fermi sea.  
We demonstrate control 
of  the magnetization reorientation 
via the photoexcitation 
frequency  and polarization direction and  
by changing the initial 
easy axis.

We  use the Hamiltonian 
$ H(t)= H_b +H_{exch}(t)+H_{L}(t)$, where 
$H_b=H_0 + H_{SO}+H_{pd}$
 describes the bands in the absence of photoexcitation
\cite{macd-anis}. 
$H_0$ describes the states in the presence  of the 
periodic lattice potential,
$H_{SO}$
is the spin--orbit interaction,
represented by on--site spin--dependent terms, while $H_{pd}$
is the mean field interaction 
of the hole spin with the {\em ground state} 
Mn spin 
${\bf S}_0$ (parallel to the easy axis)  
 \cite{rmp,chovan}.
Here we focus on the metallic regime 
(hole densities $\sim$10$^{20}$cm$^{-3}$), 
 where 
the  virtual crystal approximation
applies 
\cite{rmp}.
To describe the 
high momentum photoexcited states,
 we
diagonalized $H_b$ using the Slater-Koster $sp^3s^*$ tight--binding
Hamiltonian \cite{vogl}. 
We thus obtained conduction electron states
created by $\hat{e}_{{\bf k}n}^{\dag}$,
with energy $\varepsilon^{c}_{{\bf k} n}$,
and valence hole states  
created by 
$\hat{h}_{{\bf k} n}^{\dag}$,
with energy $\varepsilon^{v}_{{\bf k} n}$.
${\bf k}$ is the crystal momentum and  
$n$ labels the different  bands.
The 
holes experience an effective  magnetic field 
colinear to ${\bf S}_0$, acting only on $p$-orbitals, 
that lifts 
the  degeneracy of the  heavy (hh) 
and light (lh) hole  GaAs  bands 
by the magnetic exchange energy
$\Delta_{pd} = \beta c S$, where $c$ 
is the Mn  density and 
$S$  the Mn spin amplitude  \cite{rmp}. The
interaction $\beta$ 
is assumed independent of ${\bf k}$.
In II-VI semiconductors, 
a direct theory-experiment comparison
\cite{beta1} 
suggested that $\beta$ decreases along
the $\Lambda$ direction,  likely due to the ${\bf k}$--dependence 
of the hybridization with 
Mn ions
\cite{beta2}.
$H_{exch}$ is the 
 Kondo--like   interaction 
between the hole spin and the 
photoexcited deviation, $\Delta {\bf S}$,
of the Mn spin ${\bf S}$
from ${\bf S}_0$
 \cite{chovan}. 
The carrier coupling  to the optical pulse 
is characterized by the Rabi energies 
 $d_{m n {\bf k}}(t)=d_{mn{\bf k}} 
\exp{[-t^{2}/\tau_p^{2}]}$, 
where $\tau_p$=100fs is the  pulse duration 
: \cite{sbe}    
\begin{equation}
\label{HL-full} H_{L}(t)= -
\sum_{n m {\bf k}} 
d_{ mn{\bf k}}(t)  
\hat{e}^{\dag}_{{\bf k}m} \hat{h}^{\dag}_{{-\bf k}n}
+ h.c.
\end{equation} 
We consider  linearly polarized optical field 
${\bf E}$(t) 
propagating along the 
[001] crystallographic axis 
(z axis). 
The dipole matrix elements
$\mu_{m n {\bf k}}$, 
$d_{m n {\bf k}}$=${\bf \mu}_{m n {\bf k}}  {\bf E}$(t), 
were 
expressed in terms of the 
tight--binding parameters \cite{vogl} 
by considering 
the matrix elements of 
$\nabla_{{\bf k}} H_b({\bf k})$ 
\cite{voon}.

In addition to the 
photoexcited states,  
we consider the effect of the  
thermal holes.
Wang {\it et.al.} \cite{wang05} 
measured an upper bound of $\sim$200fs to the hole 
spin relaxation time in InMnAs. 
The Fermi sea spin 
relaxes on a 
time scale of 10's of fs  
due to the interplay between disorder 
and spin--orbit \cite{jung} and other \cite{wang05} interactions. 
For such fast relaxation, 
we assume to first approximation that the thermal hole spin 
adjusts adiabatically to the instantaneous ${\bf S}$(t)
\cite{chovan}
and describe the effects of the Fermi sea bath 
via its total energy $E_h({\bf S})$ \cite{macd-anis}. 
This Fermi sea populates valence 
states  close to the $\Gamma$ point. 
In view of  uncertainties such as  
 the population of impurity bands 
and the origin of strain  
\cite{welp,rmp}, 
we adopt the general form
of $E_h$ dictated by the symmetry 
\cite{macd-anis}
and extract the anisotropy parameters from the 
experimental measurements \cite{welp,merlin}: 
\begin{eqnarray} 
E_h 
 =  K_{c} (\hat{S}_x^2 
\hat{S}_y^2 + \hat{S}_x^2 
\hat{S}_z^2 + \hat{S}_y^2 
\hat{S}_z^2) 
+ K_{uz} \hat{S}_z^2
- K_{u} 
\hat{S}_x
\hat{S}_y.
\label{tot-en}
\end{eqnarray} 
The total energy $E_h$ depends on the direction of
the  magnetization unit vector 
${\bf \hat{S}}$  
\cite{macd-anis}.
In contrast to 
magnetic insulators \cite{kimel},   
the local Mn moments in (Ga,Mn)As
are 
 pure $S$=5/2 spins with angular momentum $L$=0, so
the localized electrons do not contribute to the magnetic anisotropy.  
$K_{c}$ is 
the cubic anisotropy constant, 
$K_{uz}$ is the uniaxial anisotropy constant, 
due to strain 
and shape anisotropy, and $K_u$
describes an 
in--plane uniaxial anisotropy    
observed experimentally 
\cite{welp,rmp}. 
Here we neglect for simplicity the 
light--induced temperature elevation
and assume the equilibrium values 
of the anisotropy parameters \cite{welp,merlin}. 
For sufficiently large $K_{uz}>$0,
${\bf S}_0$ lies within the $x$-$y$ plane.  
For 0$<K_u<K_{c}$, 
as is the case at low temperatures, 
there are two in-plane metastable easy axes, 
$X^-$ and $Y^+$  as in Ref.\cite{wang}, 
which point at an angle $\phi$, 
$\sin 2 \phi =  K_u/K_{c}$,  
with respect to  the x--axis [100] \cite{welp}.

With  $\sim$3eV photons, 
we excite high momentum
states along the 
\{111\} equivalent directions in the BZ, which are well separated 
in energy 
from  the thermally populated states \cite{burch}. 
We can then  
distinguish between thermal \cite{var} 
and photoexcited carrier
contributions to the 
mean field equation of motion 
of 
${\bf S}$(t)=${\bf S}_{0}$+$\Delta {\bf S}(t)$, 
\begin{equation} 
\partial_{t} {\bf S}
= - \gamma {\bf S} \times  {\bf H}^{th} 
- 
\frac{\beta}{V} \sum_{{\bf k}}{\bf S} \times \Delta {\bf s}^h_{{\bf k}}
+ \frac{\alpha}{S} {\bf S} \times 
\partial_{t} {\bf S},
\label{Mn-spin} 
\end{equation} 
where $\gamma$ is the gyromagnetic ratio, 
$\Delta {\bf s}_{{\bf k}}^h$ is the 
deviation (from its thermal value)
of the total hole spin 
\begin{equation} 
{\bf s}_{{\bf k}}^h
= 
\sum_{nn^{\prime}} 
{\bf s}^h_{{\bf k}nn^{\prime}}
\langle 
\hat{h}^{\dag}_{-{\bf k}n} \hat{h}_{-{\bf k}
  n^{\prime}}\rangle,
\label{h-spin} 
\end{equation}
$\alpha$ is the Gilbert damping coefficient
\cite{kapet}, 
and 
${\bf H}^{th} =  -\frac{\partial E_h}{\partial {\bf S}}$ 
is the thermal hole anisotropy field \cite{var}.  

 We describe   the itinerant spin and charge 
dynamics within 
the mean field approximation 
by deriving equations of motion for the 
carrier
populations 
and coherences \cite{sbe}.
The nonlinear  $e$--$h$ coherences 
are given by 
\begin{eqnarray} 
&&i \partial_{t} 
\langle \hat{h}_{-{\bf k} n} 
\hat{e}_{{\bf k} m} \rangle
= \left(\varepsilon^c_{{\bf k} m} 
+\varepsilon^v_{{\bf k} n} 
-i/T_2 \right) \langle \hat{h}_{-{\bf k} n} 
\hat{e}_{{\bf k} m} \rangle
\nonumber \\
&&
- d_{m n {\bf k}}(t) \left[1 - \langle \hat{h}^\dag_{-{\bf k}
 n} 
\hat{h}_{-{\bf k} n} \rangle 
- \langle \hat{e}^\dag_{{\bf k} m}
\hat{e}_{{\bf k} m} \rangle \right] 
\nonumber \\ && 
+ \sum_{n^\prime \ne n}
 d_{m n^\prime {\bf k}}(t) 
\langle \hat{h}^\dag_{-{\bf k} n^\prime} \hat{h}_{-{\bf k} n} \rangle 
+  \sum_{m^\prime \ne m}d_{m^\prime n {\bf k}}(t) 
\langle \hat{e}^\dag_{{\bf k} m^\prime} \hat{e}_{{\bf k} m}
\rangle 
\nonumber \\ 
&&
+ \beta c 
\sum_{n^\prime} \Delta {\bf S}(t) 
\cdot {\bf s}^h_{{\bf k}nn^\prime} \
\langle \hat{h}_{-{\bf k} n^\prime} 
\hat{e}_{{\bf k} m} \rangle.
\label{P}
\end{eqnarray} 
The nonlinear contributions to the 
above equation include 
 Phase Space Filling \cite{sbe}
(second line),
coupling to 
carrier coherences  (third line), 
and 
transient changes in the hole states due to 
interactions with 
the light--induced 
$\Delta {\bf S}$(t) 
(fourth line).  
The hole populations and inter--valence band coherences are 
determined by the equation 
 \begin{eqnarray} 
&& i \partial_t 
\langle \hat{h}^{\dag}_{-{\bf k}n} 
\hat{h}_{-{\bf k} n^{\prime}}\rangle
=
\left(\varepsilon^v_{{\bf k} n^{\prime}} 
- \varepsilon^v_{{\bf k} n} - i \Gamma^h_{nn'}\right) 
\langle \hat{h}^{\dag}_{-{\bf k}n} 
\hat{h}_{-{\bf k} n^{\prime}}\rangle 
\nonumber \\
&& + \sum_{m}
d^*_{mn {\bf k}}(t) 
\langle \hat{h}_{-{\bf k}
n^{\prime}} \hat{e}_{{\bf k} m} 
\rangle
- \sum_{m}
d_{m n^{\prime} {\bf k}}(t)  
\langle  \hat{h}_{-{\bf k}
n} \hat{e}_{{\bf k} m}
 \rangle^*
\nonumber \\
&& 
+ 
\beta c \sum_{l} \Delta {\bf S}(t) \cdot 
{\bf s}^h_{{\bf k} n^{\prime} l}
\ \langle \hat{h}^{\dag}_{-{\bf k}n} 
\hat{h}_{-{\bf k} l}
\rangle 
\nonumber \\
&& 
- \beta c \sum_{l} \Delta {\bf S}(t) \cdot 
 {\bf s}^{h*}_{{\bf k}n l} \
\langle \hat{h}^{\dag}_{-{\bf k} l} 
\hat{h}_{-{\bf k} n^{\prime}}\rangle,
 \label{dm-h} 
\end{eqnarray}
while  similar equations
are obtained for the electrons. 
We note 
that the hole populations are coupled 
to the inter--valence band coherences 
due to interactions with $\Delta {\bf S}$(t). 
$\Gamma^h_{nn^\prime}$, 
$n$$\ne$$n^\prime$, are the  dephasing 
rates  
and $\Gamma^h_{nn}$=$1/T_1$ the  population 
relaxation rate, which 
describe 
the photoexcited hole scattering  
and disorder/defect trapping effects.

\begin{figure}[t]
\vspace{0.38 in}
\centerline{
\hbox{\psfig{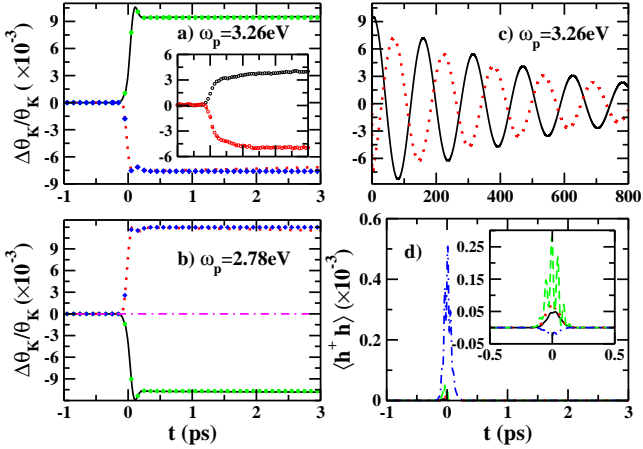}}}
\caption{(Color online)
Femtosecond Mn spin dynamics
for 
hole relaxation faster than the optical pulse.
{\bf (a)} and {\bf (b)}: 
$\Delta S_z/S$  
of the two in--plane magnetic memory states
$X^-$ (solid line) and $Y^+$ (dotted line) 
for 
two frequencies $\omega_p$ 
that excite different valence bands.
Dashed--dotted curve in Fig.\ref{Fig1}(b): 
$\omega_p$=1.6eV.
Symbols: non--thermal dynamics (${\bf H}^{th}$=0). 
Inset: Experimental $\Delta \theta_K/\theta_K$
\cite{wang} 
of $X^-$ and $Y^+$
for $\omega_p$=3.1eV. 
{\bf (c)}: 
Zero--${\bf k}$  magnon oscillations following (a).
({\bf d}): 
Populations and 
inter--valence band coherences (inset) 
corresponding to  (b). 
$\Delta_{pd}$=130meV,  $T_2$=33fs, 
$K_c$=0.014meV, 
$K_u$=$K_c$/3, 
$K_{uz}$=5$K_c$, $\alpha$=0.03, $E$=$2 \times 10^5$ V/cm
($d\sim$ 0.6--2 meV).  
}\label{Fig1}
\end{figure} 

\begin{figure}[t]
\vspace{0.38 in}
\centerline{
\hbox{\psfig{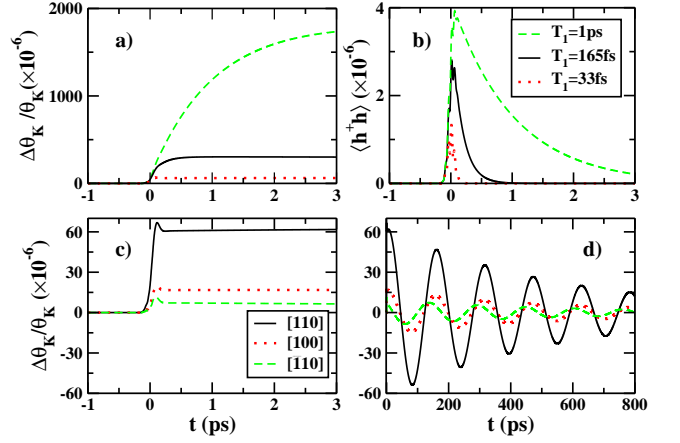}}}
\caption{(Color online)
({\bf a}): 
Femtosecond Mn spin dynamics for different $T_1$.
({\bf b})
Hole population relaxation corresponding to (a).
({\bf c}) and ({\bf d}):
Control, via the 
optical field polarization, of 
femtosecond dynamics (c) and 
Zero--momentum  magnon oscillations (d). 
$d \sim$ 0.1meV. 
Other parameters as in Fig.\ref{Fig1}(a).
}\label{Fig2}
\end{figure}

We consider ${\bf S}_0$  parallel to an
in--plane easy axis, $X^-$ or $Y^+$ \cite{wang} ($S_z$=0 initially
\cite{welp}).  
Figs.\ref{Fig1}(a) and (b)  show  the femtosecond stage of 
 the Mn spin dynamics triggered by 
a linearly polarized  pulse 
tuned close to the $\Lambda$--edge
with a fluence $\sim$7$\mu J/cm^2$
as in Ref.\cite{wang}. 
An out--of--plane  magnetization tilt 
develops on this
time scale, 
consistent with the experimental results 
for $\Delta \theta_K/\theta_K \approx \Delta S_z/S$ 
(inset), where $\theta_K$ is the Kerr rotation angle. 
This tilt practically disappears  for $\sim$1.6eV photoexcitation of 
 corresponding states close to the Fermi surface 
(dashed curve in Fig. 1(b)).  
For the anisotropy parameters of (Ga,Mn)As 
 \cite{merlin,welp},   
the thermal contribution
to Eq.(\ref{Mn-spin}), ${\bf H}^{th}$, 
plays a  minor role 
on the sub--picosecond time scale.
When the photoexcited hole 
dephasing and relaxation occurs faster 
than the pulse duration 
(Fig.\ref{Fig1}(d)), 
$\Delta {\bf S}$ 
develops {\em during the optical excitation}, due to
interactions with
the coherent $e$--$h$ pair spin. 
This initial dynamics 
is followed by a second 
temporal regime,  
Fig.\ref{Fig1}(c),  
dominated  by 
${\bf H}^{th}$  
and characterized by zero--momentum magnon oscillations 
with frequency 
\begin{equation} 
\omega^2 = \frac{ 4 \gamma^2}{S^2} 
\frac{K_{c} + K_{uz}}{K_{c}} ( K_{c}^2 - K_u^2).    
\end{equation}  
If the hole relaxation 
 is slower than the pulse duration, 
$\Delta {\bf S}$(t) 
 develops on a  time scale $\sim T_1$, 
the population relaxation time 
(Fig.\ref{Fig2}(a) and (b)).   
We conclude that the femtosecond 
magnetization reorientation 
is  governed by
the dynamics of both coherent and  nonthermal holes.

Due to its nonthermal origin, the 
femtosecond magnetization tilt can be controlled 
via the coherent photoexcitation.
First, it increases with  intensity 
(compare Figs.\ref{Fig1} and \ref{Fig2}). 
The {\em sign} of $\Delta S_z$ can  be controlled by  
tuning the photoexcitation frequency, 
 which controls 
which bands 
are excited,  
and by changing the direction of ${\bf S}_0$
(compare Figs.\ref{Fig1}(a) and (b)).  
Fig.\ref{Fig1} demonstrates femtosecond resolution of 
the easy axis direction.  
Another means of  control is 
demonstrated by 
Fig.\ref{Fig2}, which 
shows the dependence of the tilt 
(Fig.\ref{Fig2}(c)) 
and oscillations (Fig.\ref{Fig2}(d)) 
on the direction of the optical field  polarization
as the latter is rotated within the x--y plane. 
An analogous effect was observed  in 
ferrimagnetic garnets \cite{kimel}.

Similar to Ref.\cite{chovan}, 
our results  can be interpreted 
in terms of  photoexcitation
 of a pulsed hole spin component 
 $\Delta {\bf s}^{h \bot }$ 
perpendicular to ${\bf S}_0$. 
Our  mechanism should  be contrasted 
to the 
 inverse Faraday effect, which predicts that 
non--resonant excitation 
with circularly polarized light  induces an effective magnetic field 
 parallel to the direction of light propagation
in a non--absorbing 
material
\cite{farad}. 
The interpetation of  
 Ref.\cite{farad} relies on equilibrium concepts such as
free energy.
Here we develop
a  non-equilibrium 
theory that 
 treats both  interactions and 
coherent nonlinear optical effects 
similar to the Semiconductor   
Bloch equations \cite{sbe}. Our theory describes the
dynamics 
for both resonant and nonresonant 
photoexcitation
and 
takes into account the 
quantum mechanical coherence between all  bands.
We may interpret $\Delta {\bf S}$(t) as triggered 
by a {\em femtosecond} effective magnetic field pulse 
that  does {\em not} point in the 
direction of light propagation.

\begin{figure}[t]
\vspace{0.38 in}
\centerline{
\hbox{\psfig{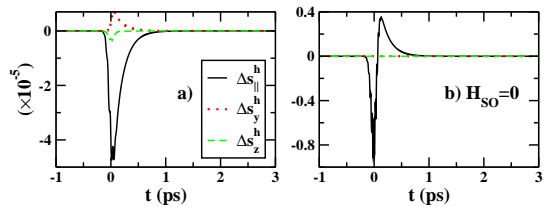}}}
\caption{(Color online)
Photoexcited total hole spin components parallel and perpendicular 
to ${\bf S}_0$:
({\bf a}) Full calculation for the parameters of 
Fig.\ref{Fig2}
({\bf b}) 
$H_{SO}$=0. 
}\label{Fig3}
\end{figure}

We now turn to the role of the interactions
$H_{pd}$ and $H_{SO}$ in 
photoexciting  
$\Delta {\bf s}^{h \bot }$. 
The hole spin, 
Eq.(\ref{h-spin}),    
 is  
determined by the density matrices $\langle \hat{h}^\dag_{-{\bf k} n} 
\hat{h}_{-{\bf k} n^{\prime}} \rangle$, 
which  can be simplified 
by expanding 
Eqs.(\ref{P}) and (\ref{dm-h})
up  to second order 
in the optical field,  
and by the spin matrix elements.
$H_{pd}$ and  $H_{SO}$  are characterized by the magnetic exchange, 
$\Delta_{pd}$,
and spin--orbit,  
$\Delta_{SO}\sim$350meV,  energies.
In samples where 
$\Delta_{pd} \gg \Delta_{SO}$,  
the hole eigenstates 
 are  spin--polarized almost parallel to
${\bf S}_0$ for all ${\bf k}$, 
i.e. ${\bf s}^{h \bot}_{{\bf k}nn }$$\approx$0.
At the same time, the coherences
$\langle \hat{h}^{\dag}_{-{\bf k}n} 
\hat{h}_{-{\bf k} n^{\prime}}\rangle$
are suppressed
when 
$\varepsilon^v_{{\bf k} n^{\prime}} 
- \varepsilon^v_{{\bf k} n}$
far exceeds the pulse frequency width,  as
for large 
$\Delta_{pd}$. 
Figure \ref{Fig3} compares
the calculated photoexcited spin components 
parallel
($\Delta {\bf s}^h_{\parallel}$)
 and perpendicular ($\Delta {\bf s}^h_y$ and 
$\Delta {\bf s}^h_z$) 
to ${\bf S}_0$ 
to the calculation with 
$H_{SO}$=0. 
$\Delta {\bf s}^{h \bot}$ is suppressed for $H_{SO}$=0
and therefore the 
light--induced femtosecond dynamics is heavily influenced by the 
(Ga,Mn)As valence bandstructure.

In the opposite limit  $\Delta_{SO}\gg\Delta_{pd}$,  
the spin of the photoexcited hole states 
is almost parallel to 
${\bf k}$, which points along the 
\{111\} equivalent directions. 
The total $\Delta {\bf s}^h$
vanishes if all symmetric directions are 
excited 
equally.
However, $H_{pd}$ and the magnetic anisotropy 
break this symmetry and introduce a preferred direction
along ${\bf S}_0$.
The band energies  now
depend on the projection of 
${\bf k}$ on ${\bf S}_0$.
The linearly polarized optical 
field introduces another preferred direction 
that changes the Rabi energies
$d_{mn{\bf k}}$. 
As a result, 
different ${\bf k}$ states are not photoexcited equally. 
Finally, for sufficiently small $\Delta_{pd}$, 
the photoexcitation 
of inter--valence band coherences becomes
significant.
 Since in (Ga,Mn)As  $\Delta_{pd} \sim$ 
150 meV is comparable to $\Delta_{SO}$, the spin
dynamics results from a competition between $H_{pd}$ and $H_{SO}$ 
that must be treated numerically.

In summary, we developed a non--equilibrium theory of ultrafast 
magnetization re--orientation
in  (Ga,Mn)As,  triggered by the 
interactions of 
photoexcited 
coherences and 
non--thermal itinerant carriers
with local Mn spins.  
We predict an initial femtosecond  regime 
of spin dynamics 
and demonstrate non--thermal  magnetization control.  
Our calculations
 explain 
recent experiments \cite{wang}.

This work was supported by the E.U. STREP program HYSWITCH, 
the U.S. National Science Foundation grant DMR0608501, 
and the U.S.
Department of Energy-Basic Energy Sciences 
under contract  DE-AC02-07CH11358.


\begin{references}


\bibitem{nagaev} 
 E. L. Nagaev, Phys. Rep. {\bf 346}, 387 (2001).


\bibitem{rmp} 
T. Jungwirth {\em et al.}, Rev. Mod. Phys. {\bf 78}, 2006.
 
\bibitem{ohno2000} 
D. Chiba {\it et.al}., Nature {\bf 455}, 515 (2008);   
J. Wang {\it et.al.}, Phys. Rev. Lett. 98, 217401 (2007).

\bibitem{kimel} 
A. V. Kimel {\it et. al}., 
Nature {\bf 435}, 655 (2005); 
F. Hansteen {\it et. al}., Phys. Rev. Lett. 
{\bf 95}, 047402 (2005).


\bibitem{wang} 
J. Wang {\it et. al}., Appl. Phys. Lett. {\bf 94}, 
021101 (2009).




\bibitem{demag}
J. Wang {\it et.al}.,  J. Phys: Cond. Matt. {\bf 18},
R501 (2006); 
L. Cywi\'{n}ski and L. J. Sham, Phys. Rev. B 
{\bf 76}, 045205 (2007);  
 E. Kojima {\it et.al}., 
 Phys. Rev. B {\bf 68}, 193203 
(2003).

\bibitem{bigot1} 
E. Beaurepaire {\it et. al.}, 
Phys. Rev. Lett. {\bf 76}, 4250 (1996).  

\bibitem{wang05}
J. Wang {\it et.al}., 
Phys. Rev. Lett. {\bf 95}, 167401 (2005).  





\bibitem{bigot-05} 
M. Vomir {\it et. al.}, 
Phys. Rev. Lett. {\bf 94}, 
237601 (2005).  


\bibitem{tolk}
J. Qi {\it et. al.},
Phys. Rev. B {\bf 79}, 085304 (2009);
Appl. Phys. Lett. {\bf 91}, 112506 (2007).  


\bibitem{merlin} D. M. Wang {\it et. al.}, Phys. Rev. B {\bf 75}, 
233308 (2007). 


\bibitem{sbe} 
W. Sch\"{a}fer and M. Wegener, 
{\em Semiconductor Optics and Transport Phenomena}
(Springer-Verlag, Berlin, 2002).


\bibitem{farad} 
P. S. Pershan {\it et. al.}, 
Phys. Rev. {\bf 143}, 574 (1966).

\bibitem{hash} 
Y. Hashimoto {\it et.al.}, Phys. Rev. Lett. {\bf 100}, 067202 (2008).

\bibitem{rozk} 
E. Rozkotov\'{a} 
{\it et. al.},
Appl. Phys. Lett. {\bf 93}, 232505 (2008). 

\bibitem{burch} 
K. S. Burch 
{\em et. al.}, Phys. Rev. B 70, 205208 (2004).



\bibitem{macd-anis} 
M. Abolfath {\em et.al.},
 Phys. Rev. B
{\bf 63}, 054418 (2001);  
 T. Dietl {\em et.al.}, Phys. Rev. B {\bf 63},
  195205 (2001).



\bibitem{chovan} 
J. Chovan, E. G. Kavousanaki, 
 and I. E. Perakis, Phys. Rev. Lett. 
 {\bf 96}, 057402 (2006); 
J. Chovan and I. E. Perakis, Phys. Rev. B {\bf 77}, 
085321 (2008).


\bibitem{vogl} 
P. Vogl {\em et.al.}, J. Phys. Chem. Solids {\bf 44}, 
365 (1983). 

\bibitem{beta1}
D. Coquillat {\em et.al.}, Phys. Rev. B {\bf 39}, 10088 (1989).

\bibitem{beta2} 
A. K. Bhattacharjee,  Phys. Rev. B {\bf 41}, 5696 (1990).


\bibitem{voon} 
L. C. Lew Yan Voon and L. R. Ram--Mohan, 
Phys. Rev. B {\bf 47}, 15500 (1993). 

\bibitem{jung} T. Jungwirth {\em et. al.}, 
Appl. Phys. Lett. {\bf 81}, 4029 (2002). 





\bibitem{welp} 
U. Welp {\em et.al.}, 
Phys. Rev. Lett. {\bf 90}, 167206 (2003); 
X. Liu {\em et. al.}, Phys. Rev. B {\bf 71}, 
035307 (2005). 


\bibitem{var} 
J. R. MacDonald, Proc. Phys. Soc. A {\bf 64}, 
968 (1951). 


\bibitem{kapet}
Y. Tserkovnyak, G. A. Fiete, and B. I. Halperin, 
Appl. Phys. Lett. {\bf 84}, 5234  (2004); 
M. D. Kapetanakis and I. E. Perakis,
Phys. Rev. Lett. {\bf 101}, 097201 (2008). 



\end{references}
\end{document}